\documentclass{mn2e}

\usepackage{amsfonts}
\usepackage{amsmath}
\usepackage{graphicx}

\title[Velocity correlations of clusters]
{Linear theory and velocity correlations of clusters}

\author[R. K. Sheth and I. Zehavi]
{Ravi K. Sheth$^{1}$ and Idit Zehavi$^{2}$\thanks{E-mail: shethrk@physics.upenn.edu (RKS); izehavi@astronomy.case.edu (IZ)}\\
$^{1}$Department of Physics \& Astronomy, University of Pennsylvania, 
      209 S. 33rd St., Philadelphia, PA 19104, USA\\
$^{2}$Department of Astronomy and CERCA, Case Western Reserve University, 
      10900 Euclid Avenue, Cleveland, OH 44106}

\newcommand{\bm}[1]{{\mbox{\boldmath $#1$}}}

\begin{document}
\pagerange{\pageref{firstpage}--\pageref{lastpage}}

\maketitle

\label{firstpage}

\begin{abstract}
Linear theory provides a reasonable description of the velocity 
correlations of biased tracers both perpendicular and parallel 
to the line of separation, provided one accounts for the fact 
that the measurement is almost always made using pair-weighted 
statistics.
This introduces an additional term which, for sufficiently biased 
tracers, may be large.  Previous work suggesting that linear theory 
was grossly in error for the components parallel to the line of 
separation ignored this term.  
\end{abstract}



\section{\protect\bigskip Introduction}
Experiments will soon measure correlations in the peculiar velocity 
field traced by massive galaxy clusters.  The linear theory 
description of these correlations as traced by dark matter particles 
whose motions are driven by gravitational instability from an 
initially Gaussian random field, has been available for about 
20 years (Gorski 1988).  
The extension of this formalism to biased tracers, within the 
framework of perturbation theory, is in Fisher (1995); 
Sheth et al. (2001b) derive similar results within the context of 
the Halo model (see Cooray \& Sheth 2002 for a review); and 
Reg\"os \& Szalay (1996) study this problem in the context of peaks 
(see Desjacques 2008 for corrections to their expressions).  
All these analyses yield consistent results.  

However, measurements of the velocity correlations of massive haloes 
in simulations have given the impression that, although biased linear 
theory provides a reasonable description of the correlations between 
the velocity components that are perpendicular to the line of 
separation $\Psi_\perp$, it is wildly discrepant for the components 
which are parallel to the line of separation $\Psi_{||}$ 
(Croft \& Efstathiou 1994; Peel 2006), except on scales larger than 
about 100 Mpc.  This discrepancy has been attributed to nonlinear 
effects, such as those described by Colberg et al. (2000) and 
Sheth \& Diaferio (2001).  

The main purpose of this short note is to show that, in fact, 
linear theory does indeed provide a good description of $\Psi_{||}$, 
provided one accounts for the fact that the measurement is actually 
pair weighted, and so the mean streaming motions may not be negligible.  
For dark matter, this mean is sufficiently small that it can be 
ignored, but ignoring the mean streaming of massive halos towards 
one another is a very bad approximation.  
We show that keeping this term in the theoretical calculation 
provides substantially better agreement with the measurements.  

Section~\ref{theory} provides the linear theory expressions 
for the velocity statistics of interest,  
Section~\ref{measurement} presents a comparison of these 
expressions with simulations, and a final section summarizes 
our results.  An Appendix provides some technical details of the 
calculation.

\section{Biased linear theory}\label{theory}
In linear theory for dark matter, the mean approach velocity 
of particle pairs along the line of separation is given by 
\begin{equation}
 {v_{12}^{\rm dm}(r)\over Hr} = -\frac{2f(\Omega)}{1+\xi(r)}
  \int {dk\over k}\, {k^3\,P(k)\over 2\pi^2}\, {j_1(kr)\over kr},
 \label{v12}
\end{equation}
where $H$ is the Hubble constant, 
$f(\Omega)\approx \Omega^{5/9}$ for flat $\Lambda$CDM models, 
\begin{equation}
 \xi(r) = \int {dk\over k}\, {k^3\,P(k)\over 2\pi^2}\, j_0(kr),
\end{equation}
and $P(k)$ is the linear theory power spectrum of the density 
fluctuation field.  
This is often called the streaming motion; the term $1+\xi$ in 
the denominator reflects the fact that the average is over all 
pairs with separation $r$.  
For what is to follow, it will be convenient to write this as 
$v_{12}^{\rm dm}(r) \equiv \langle v_{||}^{\rm dm}(r)\rangle$, 
to emphasize the fact that the motion is parallel to the separation 
vector.  
In linear theory, the mean motion perpendicular to the line of 
separation is zero:  $\langle v_\perp^{\rm dm}(r)\rangle = 0$.  

Dark matter halos are biased tracers of the underlying density 
field.  If this bias is linear, but the velocities are unbiased, 
then 
\begin{equation}
 [1 + b^2\,\xi(r)]\,\langle v_{||}(r)\rangle
    = b\, v_{12}^{\rm dm}(r)\,[1+\xi(r)],
\end{equation}
where $b$ is the linear bias factor (see Appendix).  

The linear theory velocity dispersion is 
\begin{equation}
 \sigma_{v}^2 = {(f(\Omega)H)^2\over 3}
  \int dk \,{P(k)\,W^2(kR)\over 2\pi^2} ,
 \label{sigmav}
\end{equation}
where the smoothing window is related to halo mass by setting 
$W(x) = (3/x) j_1(x)$ with $R = (3M/4\pi\bar\rho)^{1/3}$, where 
$\bar\rho$ is the comoving number density.  The factor of 3 
makes this a one-dimensional velocity dispersion.
The presence of $W$ makes the velocity dispersion decrease 
slightly with increasing halo mass (Sheth \& Diaferio 2001).  
Note that there is no additional dependence on halo bias factor.  

The two-point velocity correlation tensor is 
\begin{eqnarray}
 \Psi_{ij}({\bm r}) &=& \langle v_i({\bm x})\,v_j({\bm x+\bm r})\rangle
      \nonumber\\
  &=& \Psi_\perp(r)\,\delta_{ij} + 
       [\Psi_{||}(r) - \Psi_\perp(r)] \,\hat{\bm r}_i \hat{\bm r}_j,
\end{eqnarray}
where $r = |{\bm r}|$, $\hat{\bm r} = {\bm r}/r$, and 
$\Psi_{||}(r)$ and $\Psi_\perp(r)$ are the radial and transverse 
correlation functions (Monin \& Yaglom 1975).  
In linear theory, the velocity field is potential, 
so $\Psi_{||}(r) = {\rm d}\,r\Psi_\perp(r)/{\rm d}r$ 
(e.g. Gorski 1988).  
For Gaussian initial conditions, the linear theory 
correlation between velocity components perpendicular to the 
line of separation is 
\begin{equation}
 \Psi_\perp(r) = f(\Omega)^2\,H^2
                 \int dk\, {P(k)\,W^2(kR)\over 2\pi^2}\, {j_1(kr)\over kr},
 \label{psiperp}
\end{equation}
whereas the linear theory correlation between velocity components 
parallel to the line of separation is 
\begin{eqnarray}
 \Psi_{||}(r) &=& f(\Omega)^2\,H^2
                 \int dk\, {P(k)\,W^2(kR)\over 2\pi^2}\, j_0(kr) \nonumber\\
              && \qquad   - 2\Psi_\perp(r) 
                 - {\langle v_{||}(r)\rangle^2\, 
                    [1 + b^2\,\xi(r)]\over 4}.
 \label{psill}
\end{eqnarray}
A number of previous analyses have ignored the final term in this 
expression.  The Appendix shows why, if the velocity correlations 
are estimated using pairs of tracer particles (as is commonly done), 
it should be included.  

\section{Comparison with simulations}\label{measurement}
We compare the expressions above with measurements made in the 
Hubble Volume simulation (Evrard et al. 2002).  The background 
cosmology for this simulation is flat $\Lambda$CDM with 
$\Omega_0=0.3$, $\sigma_8=0.9$ and $H=100h$~km~s$^{-1}$ with 
$h=0.7$.  We present results for halos within a very narrow mass 
bins centered on $\log_{10} M/h^{-1}M_\odot = 14, 14.5$ and 15.  
There were 28956, 21461, and 9352 haloes in each bin.  

\begin{figure}
 \centering
 \includegraphics[width=\hsize]{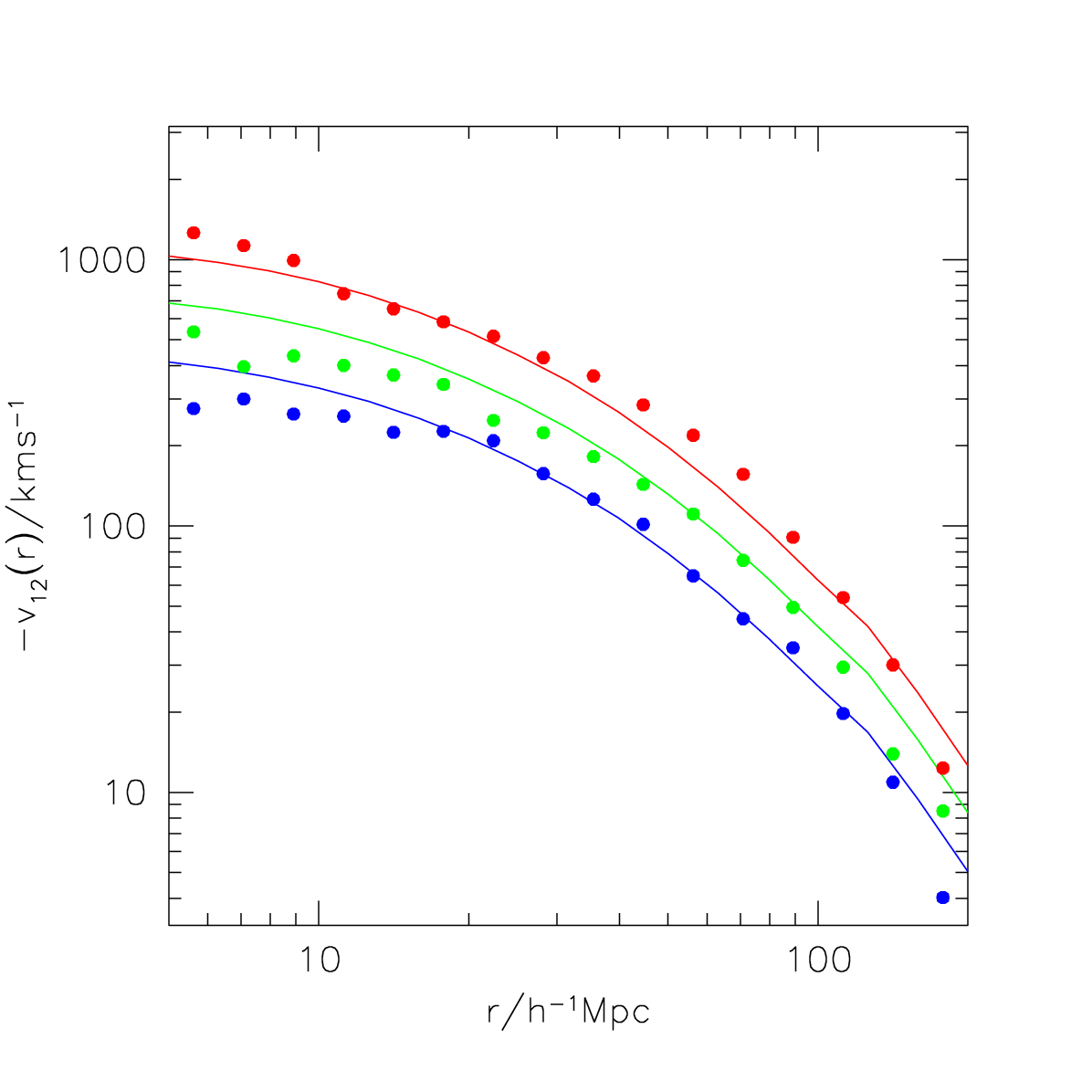}
 \caption{Mean streaming motions of halos with masses 
          $\log_{10}(M/h^{-1}M_\odot)= 14, 14.5$ and 15 (bottom to top).  
          Smooth curves show the linear bias prediction, 
          $b\,v_{12}(r)$, with $b=1.2, 2$ and 3.}
 \label{v12plot}
\end{figure}

The measured one-dimensional velocity dispersions of the halos 
are $\sigma_v =$ 313, 303 and 283~km~s$^{-1}$, whereas the values 
predicted by equation~(\ref{sigmav}) are slightly smaller: 
289, 276 and 258~km~s$^{-1}$.  This may be an indication that 
nonlinear effects of the sort discussed by Colberg et al. (2000) 
and Sheth \& Diaferio (2001) have affected halo velocities.  If 
we were to set $W\to 1$ in equation~(\ref{sigmav}) as a crude way 
of accounting for nonlinear evolution, then this would make 
$\sigma_v = 322$~km~s$^{-1}$.  

Figure~\ref{v12plot} shows the mean streaming motions of massive 
halos along the line of separation.  Notice that these velocities 
can be large:  the important point in the present context is that, 
on small scales, $v_{12}\gg \sigma_v$.  

The smooth curves show $v_{12}^{\rm dm}(r)\, [1+\xi(r)]$ of 
equation~(\ref{v12}), multiplied by bias factors of $b=1.2, 2$ and 3.  
They provide a reasonably accurate description of the measurements.  
One might have thought that a more appropriate model would not 
multiply by $1+\xi$, and would include a factor of $W^2(kR)$ in the 
integral of equation~(\ref{v12}), for the same reason that it is 
included in the expressions for $\sigma_v$, $\Psi_{||}$ and $\Psi_\perp$.  
We have found that including such smoothing produces too little 
streaming (compared to the simulations) on small scales.  
If the larger streaming on small scales is a consequence of nonlinear 
effects, then it may be that removing the smoothing term is a 
convenient way to approximately account for nonlinear evolution.  
(E.g., setting $W\to 1$ in equation~\ref{sigmav} brings $\sigma_v$ 
into slightly better agreement with the measurements.)  
Multiplying by $1+\xi$ serves a similar purpose.  

\begin{figure*}
 \centering
 \includegraphics[width=\hsize]{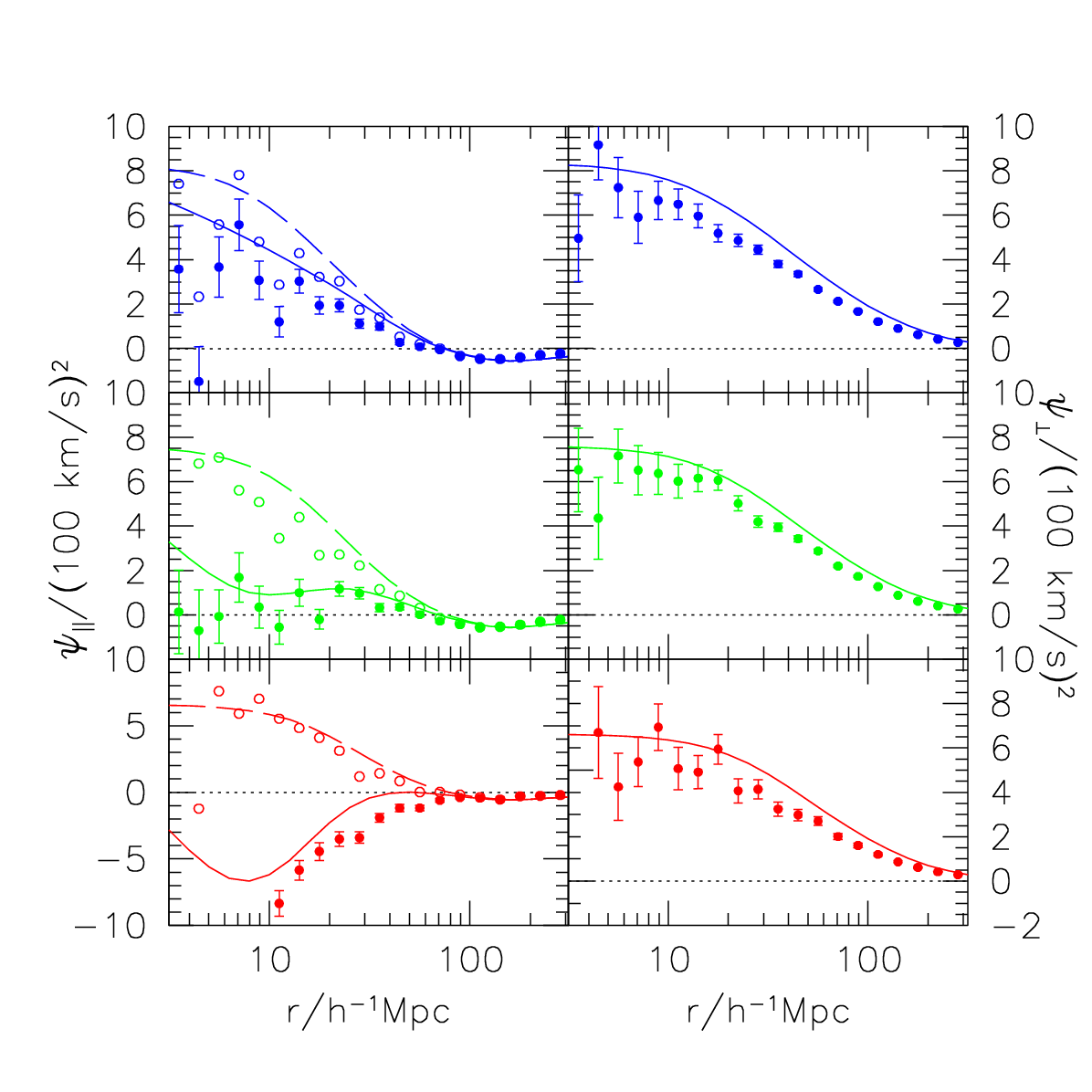}
 \caption{Velocity correlations of halos with mass 
          $\log_{10}(M/h^{-1}M_\odot)\approx 14, 14.5$ and 15 (top to bottom) 
          along (left) and perpendicular (right) to the line of separation. 
          Filled circles show the measured correlations, 
          open circles in the panels on the left show the result 
          of adding the contribution of the mean streaming motions, 
          $\langle v_{12}(r)\rangle^2/4$, to the filled circles.
          Solid curves show the linear theory predictions; dashed 
          curves in the panels on the left show the linear theory 
          prediction when the mean streaming is assumed to be negligible.}
 \label{xivkms}
\end{figure*}

Figure~\ref{xivkms} compares the predicted velocity correlations 
for these halos (curves) with the measurements (symbols).  
The solid curves show the predictions of 
equations~(\ref{psiperp}) and~(\ref{psill}).  
The dashed curves in the panels on the left show the result of 
ignoring the final term in equation~(\ref{psill}); notice how the 
dashed curves grossly overestimate the measured signal (filled symbols). 
This is the apparent failure of linear theory noted by 
Croft \& Eftathiou (1994) and Peel (2006). 
However, they are in good agreement with the measurements if, 
in the measurements, we add back the contribution from the mean 
streaming term (open symbols).  
This, and the reasonable agreement between the solid curves and 
solid symbols, indicates that linear theory is, in fact, not so 
badly flawed.  

\section{Discussion}
The mean streaming motions of massive halos can be as large as 
1000~km~s$^{-1}$ on scales of order $10h^{-1}$~Mpc.  In contrast, 
the one dimensional velocity dispersions of these halos is of 
order 300~km~s$^{-1}$.  
Nevertheless, these motions are rather well described by biased 
linear theory.  The fact that the mean streaming motions of halos 
can be substantially larger than the halo velocity dispersions 
has an important consequence:  they must not be ignored when 
making the linear theory estimate of velocity correlations 
$\Psi_{||}$ and $\Psi_\perp$ for pair-weighted statistics.  
Ignoring this contribution leads one to predict that $\Psi_{||}$ 
is positive; i.e., objects stream along with one another, whatever 
their mass.  
The measurements show that while this is true for the lower mass 
halos (they are still quite massive!), $\Psi_{||}$ is negative 
for the most massive halos.  This means that the most massive halos 
move towards, rather than along with, one another; this is 
consistent with linear theory provided one accounts for the 
fact that the measurements are pair-weighted, so one should not 
ignore the contribution of the mean streaming motions.

We note that, although they do not say so explicitly, halo model 
analyses of halo motions (Sheth et al. 2001a) have correctly 
included the effects of this mean streaming.  It is the appearance 
of more recent studies which ignore this effect (e.g. we had 
communicated the importance of this term to Peel 2006, but his 
figures do not include it) which prompted our work.  
Because our analysis shows that linear theory can be used down to 
scales which are significantly smaller than previously thought, we 
hope that our analysis will aid in the interpretation of data from 
the next generation of surveys.  

For deep surveys, the gain in accuracy is likely to be modest.  
This is because sky surveys typically estimate the correlation 
between radially projected velocities, separated by some angle 
$\theta$ on the sky.  This quantity is related to ours by 
\begin{equation}
 \Psi_{12} = \Psi_\perp \cos\theta + 
            (\Psi_{||} - \Psi_\perp)\,f(\theta,r_1,r_2),
\end{equation}
where $\Psi_{||} \equiv \sigma_v^2\,\psi_{||}$,
      $\Psi_{\perp} \equiv \sigma_v^2\,\psi_{\perp}$, 
both are evaluated at scale
 $r = \sqrt{r_1^2 + r_2^2 - 2 r_1r_2\cos\theta}$,
and 
\begin{equation}
 f(\theta,r_1,r_2) = 
   \frac{(r_1^2 + r_2^2)\cos\theta - r_1r_2(1 + \cos^2\theta)}
        {r_1^2 + r_2^2 - 2 r_1r_2\cos\theta}.
\end{equation}
This can also be written as 
\begin{equation}
 \Psi_{12}(r) = \Psi_\perp(r)\, \frac{r_1r_2}{r^2} \,\sin^2\theta + 
     \Psi_{||}(r)\,\left(\cos\theta - \frac{r_1r_2}{r^2}\,\sin^2\theta\right),
\end{equation}
showing that our extra term is most important for pairs which are 
close in angle but lie at different distances along the line of sight.  
For very deep surveys ($r_1$ and $r_2$ greater than 100 Mpc, say) the 
observable is dominated by $\Psi_\perp$ for all scales where the 
addition of our extra term is important.  Nevertheless, we 
emphasize that the term is present, and its inclusion greatly 
improves comparison with simulations, thus strengthening ones 
confidence in the validity of linear theory.  

Although accounting for mean streaming motions results in substantially 
better agreement with the simulations, there remains room for 
improvement.  Our treatment of how $\langle v_{||}(r)\rangle$ 
should depend on halo mass on small scales is rather ad hoc.  
Methods motivated by perturbation theory are beginning to provide 
more detailed prescriptions for this term (e.g. Smith et al. 2008).  
Also, in perturbation theory, the power spectrum of peculiar velocities 
is expected to be slightly suppressed relative to the linear theory 
prediction, even on relatively large scales 
(e.g. Bernardeau et al. 2002; Cooray \& Sheth 2002; 
Pueblas \& Scoccimarro 2008).
Although including this effect is beyond the scope of this work, 
we note that accounting for it will lower the theory curves slightly, 
reducing the small discrepancy in the panels on the right of 
Figure~\ref{xivkms}.  
Incorporating these refinements into our analysis is the subject of 
work in progress.

\section*{Acknowledgements}
We thank the Theoretical Astrophysics Group at Fermilab, where this 
work was started in 2000, R. Croft and A. Jaffe for helpful discussions
at that time, and the Aspen Center for Physics where this work was 
written up in 2007.  
The Hubble Volume simulation analysed in this paper was carried out 
by the Virgo Supercomputing Consortium using computers based at the 
Computing Centre of the Max-Planck Society in Garching and at the 
Edinburgh parallel Computing Centre. The data are publicly available 
at http://www.mpa-garching.mpg.de/NumCos.

\appendix
\section{Linearly biased tracers}
Dark matter halos are biased tracers of the underlying density 
field.  If this bias is linear, but the halo velocities are unbiased, 
then pair-weighted (as opposed to volume weighted) velocity 
statistics of halos may be biased relative to those of the mass, 
with the effect increasing with spatial bias.  

To see this, it is helpful write the expressions in the main text 
explicitly as averages over pairs.  
In linear theory, the mean streaming motions are 
\begin{eqnarray}
 \langle v_{||}(r)\rangle &\equiv& 
  \frac{\langle(1+b_1\delta_1)(1+b_2\delta_2)(v_{1||} - v_{2||})\rangle}
       {\langle(1+b_1\delta_1)(1+b_2\delta_2)\rangle}\nonumber\\
        &=& \frac{-b_1\, \langle\delta_1 v_{2||}\rangle 
                  + b_2\,\langle\delta_2 v_{1||}\rangle}
            {1 + b_1b_2 \langle \delta_1\delta_2\rangle}\nonumber\\ 
        &=&  -\frac{b_1+b_2}{2}\, \frac{2\langle\delta_1 v_{2||}\rangle}
                                       {1+b_1b_2\,\xi(r)}\nonumber\\ 
        &=& \frac{b_1+b_2}{2}\, v_{12}^{\rm dm}(r)\,
            \frac{1+\xi(r)}{1+b_1b_2\,\xi(r)}\nonumber\\
        &\to & b\, v_{12}^{\rm dm}(r)\,
            \frac{1+\xi(r)}{1+b^2\,\xi(r)},
\end{eqnarray}
where 
\begin{equation}
 v_{1||}-v_{2||} \equiv ({\bm v}_1-{\bm v}_2)\cdot 
   \frac{({\bm r}_1-{\bm r}_2)}{|{\bm r}_1-{\bm r}_2|}
\end{equation}
and it is understood that $r \equiv |{\bm r}_1-{\bm r}_2|$.
The factors of $1+b_1\delta_1$ and $1+b_2\delta_2$ represent the 
halo counts at positions ${\bm r}_1$ and ${\bm r}_2$, so their 
product is the pair-weight, and the averages are over all pairs 
separated by $r$.  
The algebra above follows because
 $\langle {\bm v}_1-{\bm v}_2\rangle = 0$, and because, in linear 
theory, $\langle\delta_i {\bm v}_i\rangle=0$ 
and $\langle\delta_1\delta_2 {\bm v}_1\rangle=0$.  
On large scales the difference in the pair counts is small 
$(\xi(r)\ll 1)$, and 
$v_{12} \to b\,v_{12}^{\rm dm}(r)$ where $b = (b_1+b_2)/2$.  
If the statistic is measured for a range of halo masses, then 
one simply sets both $b_1$ and $b_2$ equal to the average bias 
factor $b$ in the expressions above; this is the form we have 
used for the final expression.  
Note that $v_{12}$ is biased relative to the dark matter, even 
though the velocities themselves are explicitly unbiased.  

Similarly, in linear theory, the mean correlation of halo 
line-of-separation velocities, averaged over halo pairs, is 
\begin{eqnarray}
 \Psi_{||} &\equiv& 
  \frac{\langle(1+b_1\delta_1)(1+b_2\delta_2)\,v_{1||}\,v_{2||}\rangle}
       {\langle(1+b_1\delta_1)(1+b_2\delta_2)\rangle}
   \nonumber\\
  &=& \frac{\langle v_{1||}\,v_{2||}\rangle + 
    \langle b_1\delta_1\, b_2\delta_2\, v_{1||}\,v_{2||}\rangle}
       {1 + b_1b_2\,\langle \delta_1\delta_2\rangle}
   \nonumber\\
  &=& \frac{\langle v_{1||}\,v_{2||}\rangle + 
    b_1b_2 \langle \delta_1 \delta_2\rangle
           \langle v_{1||} v_{2||}\rangle + 
    b_1b_2 \langle \delta_1 v_{2||}\rangle
           \langle \delta_2 v_{1||}\rangle}
       {1 + b_1b_2\,\langle \delta_1\delta_2\rangle}
   \nonumber\\
  &=& \langle v_{1||}\,v_{2||}\rangle + 
    \frac{b_1b_2 \langle \delta_1 v_{2||}\rangle
           \langle \delta_2 v_{1||}\rangle}
       {1 + b_1b_2\,\langle \delta_1\delta_2\rangle}\nonumber\\
  &=& \langle v_{1||}\,v_{2||}\rangle - \frac{b_1b_2}{4}
    \frac{4\langle \delta_1 v_{2||}\rangle^2}
         {1 + b_1b_2\,\langle \delta_1\delta_2\rangle}
   \nonumber\\
  &\to & \langle v_{1||}\,v_{2||}\rangle - 
    \frac{\langle v_{||}(r)\rangle^2 \, [1 + b^2\,\xi(r)]}{4}.
\end{eqnarray}
Note that, in the second and third lines, we have assumed Gaussian 
statistics to neglect three point averages, and to write the four 
point average as a product of two-point averages.  
The final expression is what one obtains if a range of halo masses, 
with average bias factor $b$, contributes to the statistic.
On large scales, $v_{12}^{\rm dm}$ and $\xi(r)$ are both small, 
so $\Psi_{||}\approx \langle v_{1||}\,v_{2||}\rangle$.  However, 
the second term in the expression may be non-negligible when 
$b\gg 1$; neglecting it will lead to unnecessary inaccuracies.  
A similar analysis of $\Psi_\perp$ does not yield an 
extra term because $\langle\delta_2 v_{1\perp}\rangle = 0$.

\label{lastpage}

\end{document}